\documentclass[sigplan,10pt]{acmart}
\renewcommand\footnotetextcopyrightpermission[1]{}
\AtBeginDocument{%
  \fancypagestyle{standardpagestyle}{\fancyhf{}\fancyfoot[C]{\small\thepage}}%
  \fancypagestyle{firstpagestyle}{\fancyhf{}\fancyfoot[C]{\small\thepage}}%
  \pagestyle{standardpagestyle}%
}

\setcopyright{none}
\acmConference[ATC '26]{The 2026 ACM SIGOPS Annual Technical Conference}{November 2026}{Shatin, Hong Kong}
\acmYear{2026}
\acmISBN{}
\acmDOI{}

\usepackage{xcolor}
\usepackage{xspace}
\usepackage{enumitem}
\usepackage{algorithm}
\usepackage{algpseudocode}
\usepackage{booktabs}
\usepackage{multirow}
\usepackage{subcaption}
\usepackage{amsmath}
\usepackage{makecell}
\usepackage{tikz}

\newcommand{\name}[0]{StrataCL\xspace}

\newcommand{\us}[0]{{\textmu}s\xspace}

\newcommand{\codeIn}[1]{{\small\texttt{#1}}}

\newcommand{\eg}{\hbox{\emph{e.g.}}\xspace}
\newcommand{\ie}{\hbox{\emph{i.e.}}\xspace}

\newcommand{\MyPara}[1]{\noindent\textbf{\textit{#1}}~}

\definecolor{ForestGreen}{RGB}{34,139,34}
\definecolor{DarkGreen}{rgb}{0.0, 0.5, 0.0} 

\let\origref\ref
\renewcommand{\ref}[1]{\textcolor{DarkGreen}{\origref{#1}}}

\newcommand{\secref}[1]{\S\textcolor{DarkGreen}{\ref{#1}}}
\newcommand{\figref}[1]{\textcolor{DarkGreen}{\ref{#1}}}

\newcommand*\blackcircled[1]{%
  \tikz[baseline=(char.base)]{
    \node[shape=circle,draw,fill=black,text=white,inner sep=1pt] (char) {#1};
  }%
}

\title{\name: Fabric-Native Communication Library for Production Supernodes}

\author{
Tiancheng Hu$^{1,6}$,
Jin Qin$^{2,3}$,
Yuzheng Wang$^{1,6}$,
Ke Liu$^{2,3}$,
TangShengsheng Li$^{4}$,
Sheng Wang$^{4}$,
Zhongzhe Hu$^{4}$,
Tianlun Hu$^{4}$,
Wei Wang$^{4}$,
Lijun Li$^{4}$,
Jingbin Zhou$^{4}$,
Xiaoming Bao$^{4}$,
Hongwei Sun$^{4}$,
Jieru Zhao$^{5}$,
Huimin Cui$^{2,3}$,
Tao Xie$^{1,6}$,
Chenxi Wang$^{2,3}$
\\[0.6em]
$^{1}$Peking University
\qquad
$^{2}$SKLP, Institute of Computing Technology, CAS
\\
$^{3}$University of Chinese Academy of Sciences
\qquad
$^{4}$Huawei Technologies Company Ltd.
\\
$^{5}$Shanghai Jiao Tong University
\\
$^{6}$Beijing Tongming Lake Information
Technology Application Innovation Center, China
}

\begin{document}

\begin{abstract}
Modern distributed AI workloads run across hundreds of accelerators, making communication a major bottleneck. Existing communication libraries remain largely buffer-centric because user and communication buffers are managed separately, causing redundant data copies or costly user-buffer registration. This paper presents \name, a zero-redundancy and fabric-native communication library for production supernodes. \name introduces registration-on-allocation to realize user-buffer direct communication, and designs communication operators with workload-balanced NPU-core partitioning and NPU-driven SDMA offloading to exploit supernode architecture features. On the Huawei CloudMatrix384, \name improves collective bus bandwidth by up to $1.6\times$ and improves MoE dispatch/combine bus bandwidth by up to $1.4\times$. Across three production workloads, \name improves LLM inference throughput by $1.9\times$, reduces P99 TTFT by $2.2\times$, and reduces LLM and Recsys training iteration time by $1.4\times$ and $1.3\times$, respectively.
\end{abstract}

\maketitle

\section{Introduction}
\label{sec:introduction}

Modern large-scale production applications increasingly rely on distributed execution across hundreds of accelerators, making inter-device communication a major performance bottleneck~\cite{commnucation_study,megascale-infer}. For example, communication operators account for approximately 10\%--40\% of end-to-end time in distributed large language model (LLM) inference~\cite{TokenWeave,nvidia-jax-xla-25} and 30\%--45\% in training~\cite{megascale-MoE,T3}. As model size and cluster scale continue to grow, the communication-to-computation ratio further increases and may exceed 50\% when compute capability scales faster than interconnect bandwidth~\cite{tale-of-two-cs}.

Common communication libraries, such as NCCL~\cite{nccl}, RCCL~\cite{rccl}, and HCCL~\cite{hccl}, are the backbone of large-scale training and serving on GPU~\cite{NVIDIA,AMD} and NPU~\cite{Ascend} clusters. However, their conventional paths are largely \emph{buffer-centric}~\cite{NCCLX}. User buffers are allocated by application frameworks (\eg, PyTorch~\cite{PyTorch}, SGLang~\cite{sglang}), while communication buffers are separately managed by the communication runtime. This allocator-level separation prevents the communication runtime from directly operating on user buffers, as they are not registered as remotely accessible. As a result, data incurs redundant staged copies through internal communication buffers, increasing latency and consuming extra HBM, as demonstrated by previous work~\cite{NCCLX, swiftep}. Recent libraries mitigate this overhead through user-buffer direct communication, such as NCCL User Buffer Registration~\cite{nccl-ubr} and HCCL-zerocopy~\cite{hccl-zerocopy}.

On traditional clusters built over scale-out RDMA fabrics, \eg, InfiniBand~\cite{IB} and RoCE~\cite{RoCE}, redundant data movement remains difficult to eliminate, and deploying user-buffer registration is challenging. RDMA-based memory registration is costly~\cite{srnic-nsdi23,clio,mitosis}, which needs to pin the GPU pages, establish DMA mappings through PCIe BAR, create metadata such as memory regions, and exchange the information across ranks. These steps introduce non-negligible overhead, which can reach several milliseconds~\cite{gpudirect-rdma} and may offset the benefit of user-buffer direct communication, especially under dynamic allocation patterns such as MoE serving.


\MyPara{Opportunities in supernode architectures.}
Supernodes, such as GB200 NVL72~\cite{gb200nvl72} and CloudMatrix384~\cite{cm384}, are emerging server architectures that tightly integrate hundreds of accelerators through scale-up fabrics, such as NVLink~\cite{nvlink} and UB~\cite{ub-mesh}. This paper focuses on Huawei CloudMatrix384 (CM384)~\cite{xdeepserve,cm384} as a representative platform. CM384 is built around the Unified Bus (UB), a scale-up fabric that connects 384 Ascend NPUs within a single supernode (architecture details in \secref{sec:bg-supernode}). The following discussion highlights the opportunities that such supernode architectures create for communication-library design.

\emph{High-bandwidth and low-latency inter-NPU communication.}
UB provides nearly 400~GB/s bandwidth and nanosecond-scale remote HBM access latency, substantially narrowing the gap between remote and local HBM access. Within a supernode, UB data transfer is fast enough that copies between user buffers and internal communication buffers can account for a significant fraction of communication latency, making user-buffer direct communication especially beneficial (\secref{sec:bg-ubr}).

\emph{Global unified physical address space.}
CM384 maps the HBM physical address ranges of all NPUs into a unified UB address space. A memory region can therefore become remotely accessible by mapping a local virtual address to the corresponding remote NPU HBM physical address, which fundamentally simplifies remote-access memory registration and makes it much more lightweight (\secref{sec:bg-register}).

\MyPara{Challenges.}
Despite these opportunities, fully exploiting CM384's architectural features for communication optimization faces two critical challenges:

\emph{How to remove user-buffer registration overhead from the communication critical path.}
Although the global unified physical address space makes remote-access memory registration much more lightweight (\ie, microsecond-scale), just-in-time registration still adds non-negligible latency to communication, especially in large-scale scenarios due to poor scalability. A naive solution is to pre-register a sufficiently large memory pool and require the application framework to allocate user buffers from this pool. While this removes the registration overhead from the critical path, it leads to severe memory waste and intrusive framework modifications (\secref{sec:mot-register}).

\emph{How to design fabric-native communication operators for supernodes.}
With UB's high bandwidth and low latency, data transfer latency becomes less dominant in communication, while synchronization overhead becomes increasingly visible, especially for small payloads. Moreover, UB's hierarchical topology introduces non-uniform memory access. Under topology-unaware workload partitioning, NPU cores assigned to slower or heavier peer transfers may finish much later than others, creating a long-tail problem. Finally, saturating UB bandwidth requires many NPU cores to issue remote-memory operations concurrently, which can contend with overlapped compute kernels and reduce the benefit of compute--communication overlap (\secref{sec:mot-operator}). 

\MyPara{\name.} 
This paper proposes \name, a zero-redundancy and fabric-native communication library for production supernodes. \name eliminates redundant data copies along the communication path and designs communication operators considering supernode architecture features, directly addressing the two challenges above.

For the first challenge, \name proposes \emph{registration-on-allocation}, based on the key insight that a long interval typically separates a buffer's physical memory allocation from its first use in communication. For each physical allocation, \name intercepts memory APIs and asynchronously registers the region in the background, deferring only a lightweight completion check until a communication operator first accesses it. \name further uses shadow virtual addressing, assigning each NPU a disjoint virtual-address range and mirroring allocated buffers at the same virtual address on peer NPUs. This allows communication operators to issue remote access without peer-to-peer address translation. To support VMM API-based allocators, \name incrementally updates shadow mappings for remapped regions and uses a fast broadcast path to exchange registration metadata (\secref{sec:memory_management}).

For the second challenge, \name builds an efficient \emph{full-mesh} programming abstraction to avoid the frequent synchronization overhead of multi-step algorithms using remote load/store instructions. To address core-level long tails caused by non-uniform memory access of UB, \name proposes a \emph{workload-balanced NPU-core partitioning} to avoid the long-tail problem.
Finally, under compute--communication overlap, saturating UB bandwidth requires many NPU cores to concurrently issue remote-memory access instructions, which can contend with overlapped compute kernels. \name addresses this with NPU-driven SDMA offloading, where NPU cores only submit DMA descriptors and the SDMA engine asynchronously performs data movement, quickly releasing NPU cores back to overlapped computation (\secref{sec:supernode_operator}).

\name is evaluated on a CM384 supernode using up to 512 NPU dies and three real-world production applications. The results show that \name improves both operator-level communication efficiency and end-to-end application performance. In microbenchmarks, \name achieves up to $1.6\times$ higher bus bandwidth for collective communication and up to $1.4\times$ higher bus bandwidth for MoE dispatch/combine compared with state-of-the-art production communication libraries. In end-to-end evaluation, \name improves LLM inference throughput by $1.9\times$ and reduces P99 TTFT by $2.2\times$. For training workloads, \name accelerates LLM training and recommendation-model training by up to $1.4\times$ and $1.3\times$.


This paper makes the following key contributions:
\begin{itemize}
    \item It proposes \emph{registration-on-allocation}, enabling transparent user-buffer direct communication while keeping registration off the critical path, remaining compatible with VMM API-based allocators, and demonstrating portability on NVIDIA GPUs.
    
   \item It designs fabric-native communication operators for supernodes, combining a full-mesh programming abstraction, workload-balanced NPU-core partitioning, and NPU-driven SDMA offloading to reduce synchronization overhead, core-level long tails, and NPU-core contention under compute--communication overlap.
    
    \item It integrates \name into production frameworks and demonstrates significant end-to-end speedups. The source code will be released upon publication.
\end{itemize}

\begin{figure}[t]
\centering
\includegraphics[width=\linewidth]{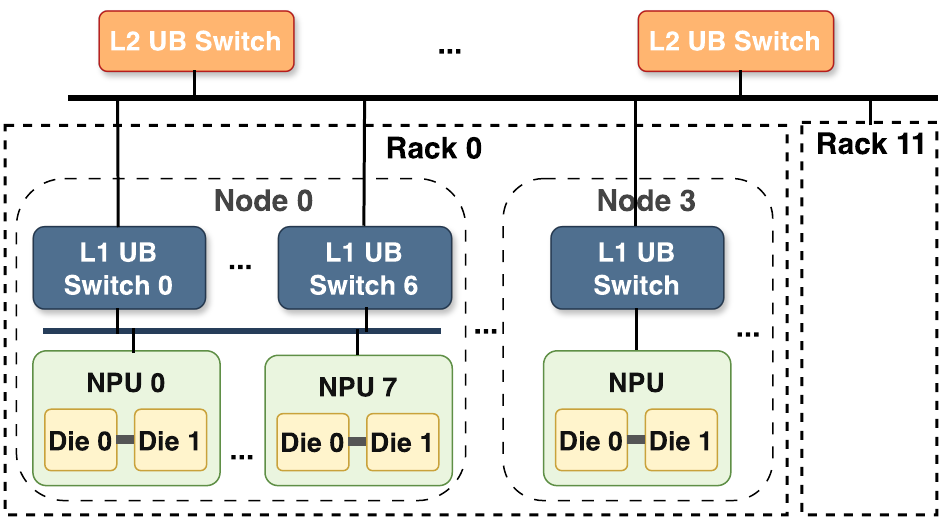}
\caption{Huawei CloudMatrix384 supernode topology.}
\label{fig:bg-supernode-arch}
\end{figure}

\section{Background and Observations}
\label{sec:background}

\subsection{Supernode Architectures}
\label{sec:bg-supernode}

CM384 integrates 384 Ascend 910C NPUs~\cite{Ascend} into a supernode interconnected by UB. As shown in Figure~\figref{fig:bg-supernode-arch}, a CM384 contains 12 compute racks, each with four nodes and eight NPUs per node. CM384 uses a two-level UB switching hierarchy in which L1 UB switches connect the eight NPUs within a node, while L2 UB switches aggregate L1 switches across nodes for inter-node connectivity.

\begin{figure}[t]
\centering
\includegraphics[width=0.8\linewidth]{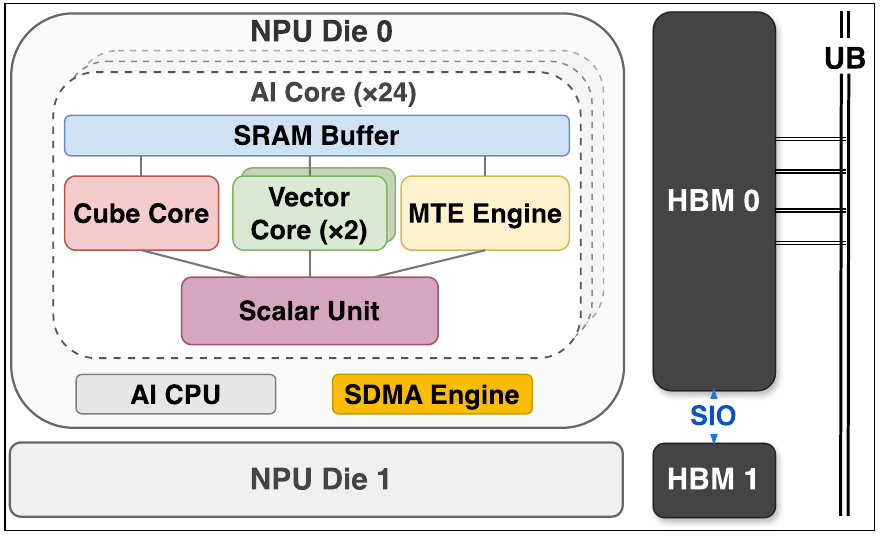}
\caption{Simplified architecture of an Ascend 910C NPU. Each NPU contains two dies, each with 24 AI Cores; each AI Core integrates one Cube Core and two Vector Cores.}
\label{fig:bg-ascend-910c-arch}
\end{figure}

Each Ascend 910C NPU is a dual-die package connected by a high-bandwidth SIO fabric with up to 540~GB/s aggregate bandwidth. As shown in Figure~\figref{fig:bg-ascend-910c-arch}, each die contains 24 AI Cores. Each AI Core integrates one Cube Unit (AIC) for matrix computation and two Vector Cores (AIV) for vector processing, coordinated by a scalar unit. Since communication kernels mainly run on AIV cores, this paper uses \emph{NPU core} to refer to an AIV core. Each die has 64~GB HBM with 1.6~TB/s bandwidth. Each AI core includes a KB-level unified buffer and Memory Transfer Engines (MTEs) for data movement between HBM and on-chip SRAM buffers. An SDMA engine further supports inter-NPU data movement, typically launched through host APIs.

\subsection{User-Buffer Direct Communication}
\label{sec:bg-ubr}

\begin{figure}[]
\centering
\begin{subfigure}[]{0.48\columnwidth}
  \centering
  \includegraphics[width=\linewidth]{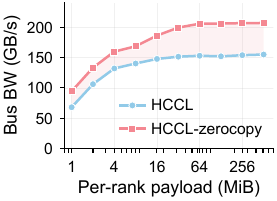}
  \phantomsubcaption
  \label{fig:mot-allgather-data}
\end{subfigure}\hfill
\begin{subfigure}[]{0.48\columnwidth}
  \centering
  \includegraphics[width=\linewidth]{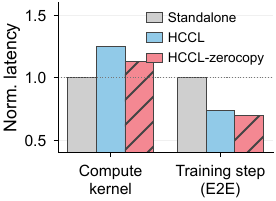}
  \phantomsubcaption
  \label{fig:mot-bandwidth-overlap}
\end{subfigure}
\caption{ (a) Bus bandwidth of  HCCL vs.\ HCCL-zerocopy AllGather at 32 ranks. (b) Normalized compute-kernel and training step latency in overlap scenarios.}
\label{fig:mot-staging-cost}
\end{figure}

Compared with RDMA, which typically provides only tens of GB/s of effective bandwidth, UB offers substantially higher bandwidth and narrows the gap between inter-NPU communication and local HBM access. This makes user-buffer direct communication especially beneficial on supernodes. Figure~\figref{fig:mot-allgather-data} reports the AllGather bus bandwidth~\cite{bus_bandwidth_NCCL} of HCCL and HCCL-zerocopy at 32 ranks. HCCL-zerocopy improves bus bandwidth by more than 30\% for all payloads above 8~MiB by eliminating redundant data movement on the staging path. For an AllGather across $N$ ranks with $M$ bytes per rank, HCCL copies $M$ bytes from each user input buffer into an internal communication buffer and then copies the gathered $N\!\cdot\!M$ bytes back to user output buffers, adding $(1{+}N)\!\cdot\!M$ bytes of staging traffic. HCCL-zerocopy bypasses these endpoint copies by communicating directly over user buffers. Although chunked-pipelined execution can overlap part of the staging cost with network transfer, it cannot fully hide the extra copy overhead, and the pipeline still occupies NPU cores and contends with concurrent compute kernels under compute--communication overlap~\cite{NCCLX,swiftep}.

User-buffer direct communication also reduces bandwidth interference. In overlap-heavy training pipelines such as Fully Sharded Data Parallel (FSDP)~\cite{fsdp}, parameter AllGather prefetches run concurrently with compute kernels, while staged copies consume HBM bandwidth that would otherwise be available to computation. Figure~\figref{fig:mot-bandwidth-overlap} quantifies this effect in a DeepSeek V3.2 671B~\cite{deepseek_v3_2} training deployment on 512 NPU dies, with detailed settings in \secref{sec:eval-end2end}. HCCL slows the concurrent compute kernel by 25\%, whereas HCCL-zerocopy reduces the slowdown to 13\%. This lower interference directly improves training performance. Compared with a no-overlap serial baseline, overlapped execution reduces training step time by 26\% with HCCL and by 31\% with HCCL-zerocopy, yielding an additional 6\% end-to-end speedup from eliminating staging-induced contention.

\subsection{Inter-NPU Memory Mapping}
\label{sec:bg-register}

\begin{figure}[]
\centering
\begin{subfigure}[]{0.53\columnwidth}
  \centering
  \includegraphics[width=\linewidth]{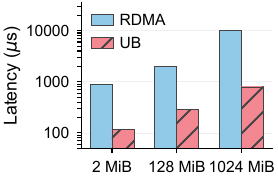}
  \phantomsubcaption
  \label{fig:bg-register-total}
\end{subfigure}\hfill
\begin{subfigure}[]{0.44\columnwidth}
  \centering
  \includegraphics[width=\linewidth]{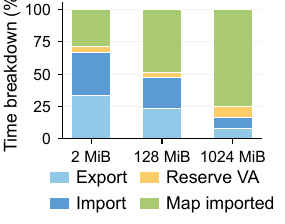}
  \phantomsubcaption
  \label{fig:bg-register-breakdown}
\end{subfigure}
\caption{Inter-NPU memory registration on supernode. (a)
Registration latency comparison of RDMA and UB. (b) Time breakdown of UB registration, normalized to the total.}
\label{fig:bg-register-overhead}
\end{figure}

The \emph{global unified physical address space} of UB makes inter-NPU memory access structurally simpler than on RDMA. Accessing a remote buffer over RDMA requires the target endpoint to register the buffer and send the remote key and remote virtual address to peer RNICs. In contrast, UB exposes remote HBM through a unified physical address space. A local virtual address can be mapped to remote HBM physical pages and then accessed like local HBM. Figure~\figref{fig:bg-register-total} reports memory registration latency under three payload sizes. UB registration is consistently an order of magnitude faster than RDMA registration, achieving a $9\times$ speedup on average. 

UB registration consists of four logical steps. The remote rank exports a physical-memory handle, and the local rank imports the handle, reserves a local virtual address, and maps the reserved VA to the remote rank's HBM. Figure~\figref{fig:bg-register-breakdown} shows the latency breakdown of these steps. For smaller payloads, handle export and import can cause significant delays. As the payload grows, the mapping step becomes dominant because NPU page-table updates scale with the registered size.

\section{Motivation}
\label{sec:motivation}


\begin{figure}[]
\centering
\begin{subfigure}[]{0.49\columnwidth}
  \centering
  \includegraphics[width=\linewidth]{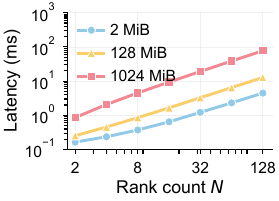}
  \phantomsubcaption
  \label{fig:mot-register-scaling}
\end{subfigure}\hfill
\begin{subfigure}[]{0.49\columnwidth}
  \centering
  \includegraphics[width=\linewidth]{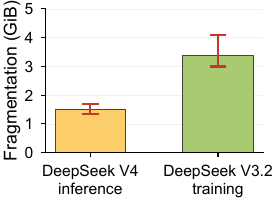}
  \phantomsubcaption
  \label{fig:mot-register-frag}
\end{subfigure}
\caption{(a) User-buffer registration overhead varying with the number of ranks. (b) HBM fragmentation on CM384 when PyTorch's expandable-segment allocator is disabled.}
\label{fig:mot-register-overhead}
\end{figure}

\subsection{User-buffer Registration}
\label{sec:mot-register}

\MyPara{Costly and poorly scalable just-in-time registration.}
Although UB reduces inter-NPU memory registration to the microsecond scale, this overhead remains non-negligible compared with communication operators, whose latency typically ranges from hundreds of microseconds to tens of milliseconds. More importantly, just-in-time registration scales poorly with rank count since each mapping operation acquires a write-side page-table lock, and per-peer mappings on the same rank must be serialized. As a result, registering a full $N$-rank communication group incurs approximately $O(N)$ latency per rank and grows nearly linearly with rank count, as shown in Figure~\figref{fig:mot-register-scaling}. Caching registered buffers can amortize this cost for static and repetitive communication patterns, but it is much less effective under dynamic memory allocation, where tensor shapes change across invocations. A representative example is MoE inference, whose dispatch buffer sizes vary across batches due to dynamic token routing~\cite{deepep}. In such cases, just-in-time registration repeatedly places the registration overhead on the communication critical path, weakening or even offsetting the benefit of user-buffer direct communication.

\MyPara{Pool-based approaches fail to achieve both HBM efficiency and framework compatibility.}
A simple workaround is to pre-register a communication memory pool~\cite{swiftep}, which removes registration from the critical path but turns registration overhead into an HBM capacity tax. Since this pool is dedicated to communication buffers, its capacity is isolated from the application framework allocator, reducing the maximum supported batch size. Some approaches~\cite{NCCLX} allow the framework and communication library to share a registered pool, but they rely on static virtual-to-physical mappings, which conflict with modern on-demand VMM API-based memory management such as PyTorch's expandable-segment allocator~\cite{pytorch-expandable}. Figure~\figref{fig:mot-register-frag} quantifies the memory fragmentation\footnote{Memory fragmentation here refers to the sum of the available memory blocks in PyTorch under the current maximum supported batch size.} on CM384 when expandable segments are disabled. DeepSeek V4 Flash~\cite{deepseek_v4} inference loses 1--2~GiB of usable memory per NPU die, while DeepSeek V3.2 671B training loses 3--4~GiB, which directly reduces the supported batch size and leads to end-to-end throughput degradation.

\begin{figure}[]
\centering
\begin{subfigure}[]{0.49\columnwidth}
  \centering
  \includegraphics[width=\linewidth]{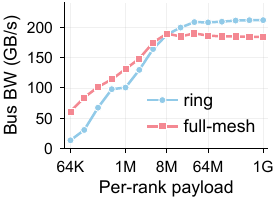}
  \phantomsubcaption
  \label{fig:mot-ring-vs-mesh}
\end{subfigure}\hfill
\begin{subfigure}[]{0.49\columnwidth}
  \centering
  \includegraphics[width=\linewidth]{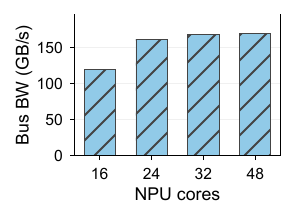}
  \phantomsubcaption
  \label{fig:mot-mesh-aiv}
\end{subfigure}
\caption{(a) Bus bandwidth of ring vs.\ full-mesh AllGather. (b) Achievable UB bus bandwidth of current NPU cores.}
\label{fig:mot-operator}
\end{figure}

\begin{table}[]
    \centering
    \footnotesize
    \caption{Access latency and single-direction bandwidth of two NPUs across different access relative locations.}
    \label{tab:bg-access-perf}
    \begin{tabular}{ccc}
        \toprule
        \textbf{Relative location}           & \textbf{Latency (\us)} & \textbf{Bandwidth (GB/s)} \\
        \midrule
        Die-to-Die       & 0.2                    & 210                      \\
        Intra-node        & 0.7                    & 170                       \\
        Inter-node        & 2.1                    & 150                       \\
        \bottomrule
    \end{tabular}
\end{table}

\subsection{Communication Operator Execution}
\label{sec:mot-operator}

\MyPara{Synchronization overhead becomes a critical bottleneck.}\\
Algorithms such as ring~\cite{allreduce}, PAT~\cite{pat}, and recursive halving--doubling~\cite{reducescatter} are effective in RDMA scale-out clusters because they reduce instantaneous fanout, mitigate network contention, and avoid excessive traffic on slow cross-node paths. However, these algorithms rely on multiple interdependent communication steps, whose cumulative synchronization overhead becomes a critical bottleneck in supernode fabrics, especially for small-to-medium messages. In contrast, full-mesh communication completes in a single logical step by issuing concurrent memory instructions.  As shown in Figure~\figref{fig:mot-ring-vs-mesh}, full-mesh outperforms ring across the small-to-medium range, on average over $2\times$ faster below 1~MiB and up to $4.5\times$ at 64~KiB, and remains ahead through 8~MiB. The latency breakdown shows that synchronization accounts for more than 50\% of the ring's end-to-end time at small payloads and reaches ${\sim}77\%$ at 64~KiB. Ring becomes preferable beyond 16~MiB, when lower cross-node traffic and fanout contention outweigh the synchronization cost.

\MyPara{Non-Uniform Memory Access (NUMA) creates core-level long tails.}
Although the supernode exposes a unified global address space, its two-level switching topology makes memory-access performance depend on the relative location of source and destination NPUs. As shown in Table~\figref{tab:bg-access-perf}, die-to-die access within the same NPU reaches 210~GB/s at 0.2~\us, while intra-node remote HBM access drops to 170~GB/s at 0.7~\us, and inter-node access further drops to 150~GB/s at 2.1~\us. Compared with die-to-die access, inter-node access has roughly $10\times$ higher latency and 29\% lower bandwidth. This non-uniform memory access cost makes NPU cores assigned to slower or heavier peer transfers finish much later than others, creating a core-level long tail problem, especially for high-concurrency full-mesh operators.

\MyPara{NPU-core contention limits overlap.}
To saturate UB bandwidth, an NPU must issue many concurrent remote-memory operations, such as remote loads/stores or peer memory copies, which requires many NPU cores to participate in data movement. As shown in Figure~\figref{fig:mot-mesh-aiv}, bus bandwidth reaches 95\% of peak only when 24 NPU cores are used, consuming half of the 48 NPU cores on each NPU die. This substantially reduces the NPU-core budget available for concurrent computation kernels and can cause severe NPU-core contention in compute--communication overlap scenarios, such as FSDP parameter prefetch~\cite{fsdp} in LLM training and Two-micro-Batch Overlapping~\cite{profile-data} in LLM inference.


\begin{figure}[t]
\centering
\includegraphics[width=.45\textwidth]{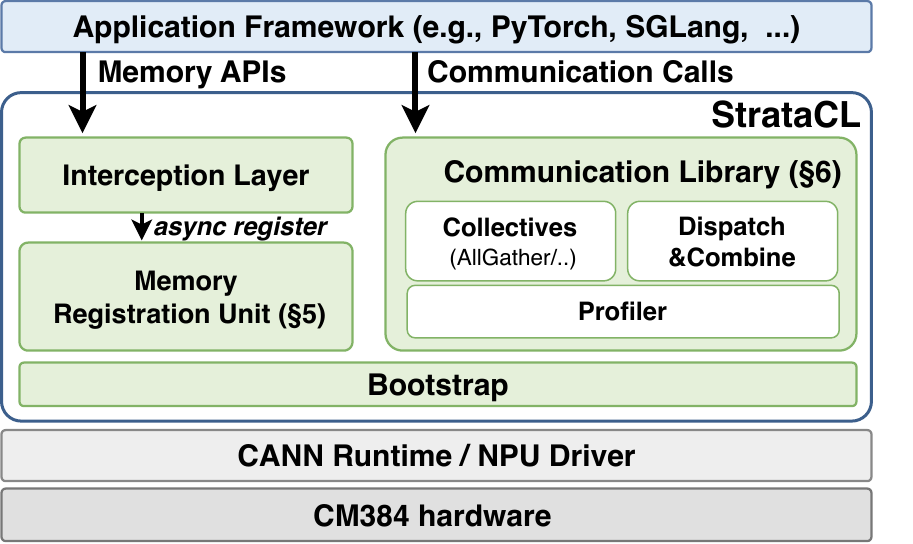}
\caption{\name{} design overview.}
\label{fig:design-overview}
\end{figure}

\section{Design Overview}
\label{sec:design_overview}

As shown in Figure~\figref{fig:design-overview}, \name sits between the application framework and the CM384 runtime stack and consists of three main modules. First, the \emph{memory registration unit} transparently makes framework-allocated user buffers remote-accessible. Through the interception layer, \name hooks application memory APIs and triggers asynchronous memory registration in the background. Second, the \emph{communication library} provides both common collective operators and workload-specific operators. It also integrates a fine-grained NPU-core profiler to identify operator bottlenecks and guide optimization. Finally, the \emph{bootstrap} component initializes communicator groups, establishes metadata-exchange channels, and assigns each NPU a disjoint virtual-address range.

\section{Memory Registration}
\label{sec:memory_management}


\subsection{Registration-on-allocation}
\label{sec:reg-on-alloc}

The key insight behind \name is that a buffer's physical memory allocation is usually separated from its first communication use by a long interval, \eg, at least 2.6~s in LLM inference (\secref{sec:ablation-gap}). Modern framework allocators reserve large virtual-address segments at startup and reuse them across iterations~\cite{vllm,sglang,pytorch_cca}. Therefore, common-case tensor allocation typically obtains a virtual address from pre-mapped memory segments, such as those managed by the PyTorch caching allocator~\cite{pytorch_cca}. \name exploits this allocation-to-communication gap with an asynchronous \emph{registration-on-allocation} mechanism. Instead of registering a buffer when it is accessed by a communication operator, \name registers the underlying physical allocation for remote access in the background immediately after allocation. This keeps remote-access memory registration off the critical path while remaining transparent to the application framework.

\begin{figure}[t]
\centering
\includegraphics[width=0.47\textwidth]{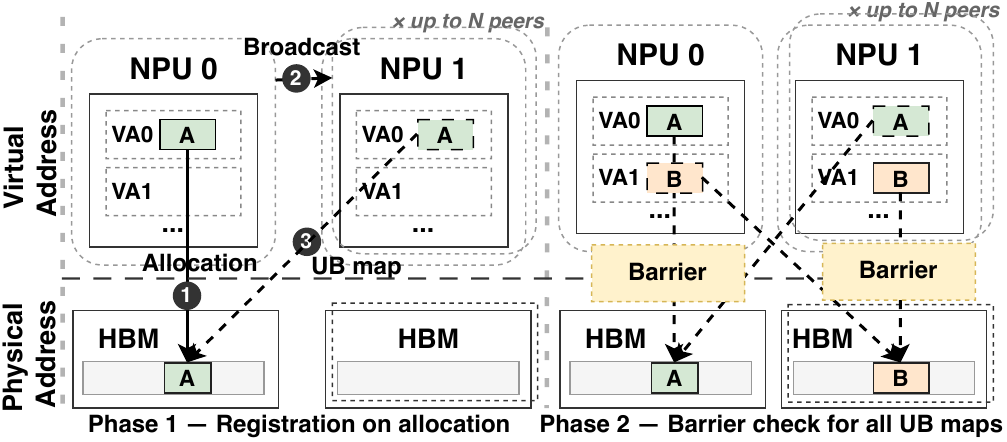}
\caption{Registration workflow in \name{}. When a physical allocation is intercepted, an asynchronous registration starts (Phase 1). When a communication operator first accesses the region, \name performs a readiness barrier to ensure that the corresponding UB mappings have completed (Phase~2).}
\label{fig:mem-reg-flow}
\end{figure}

Figure~\figref{fig:mem-reg-flow} illustrates the workflow. When the application on NPU~0 allocates buffer~$A$, \name makes $A$ remotely accessible in the background without blocking the application. \blackcircled{1} The application allocates a virtual address slot and maps it to the corresponding local HBM physical pages, following the normal allocation path. \blackcircled{2} After the allocation returns, the memory-registration unit asynchronously exports $A$'s physical-memory handle and broadcasts the required metadata to peer NPUs, including the owner NPU, virtual address, size, and physical mapping information. \blackcircled{3} Upon receiving the metadata, each peer NPU maps the same virtual address to NPU~0's remote HBM physical pages through UB (Phase~1). 
To ensure correctness, \name defers a readiness check until a communication operator first touches the region (Phase~2). This check is purely local, where each NPU only verifies that its own UB mappings for the region have completed, without cross-NPU synchronization, incurring negligible overhead.

\subsection{Shadow Virtual Addressing}
\label{sec:shadow-va}

A physical memory may be mapped by different virtual addresses on different NPUs. This forces communication operators to translate peer-buffer addresses before remote accesses and requires each NPU to maintain address-translation metadata, which becomes increasingly expensive at scale.

\name eliminates this overhead with \emph{shadow virtual addressing}, which makes a buffer visible at the same virtual address on all NPUs. The key idea is to \emph{decouple virtual-address planning from physical-memory ownership}. During initialization, \name assigns each NPU a disjoint virtual-address range, so the ranges of different NPUs never overlap. Since the virtual-address space is much larger than the physical HBM capacity, reserving these disjoint ranges introduces negligible address-space pressure. When NPU~$i$ allocates a buffer at virtual address $v$, \name performs a shadow mapping on every peer NPU. Each peer reserves the same virtual address $v$ in its own address space and maps it to the physical HBM pages of NPU~$i$ through UB's global unified physical address space. After this mapping, all NPUs can access the buffer using the identical virtual address $v$. As a result, communication operators can issue remote accesses without peer-to-peer address translation, simplifying the implementation and reducing overhead.

\subsection{Integration with VMM API-based allocator}
\label{sec:vmm}

The mechanisms above assume that a buffer's virtual-to-physical mapping remains fixed after allocation, so registration can complete before the buffer is first used. VMM API-based allocators, such as PyTorch's expandable-segment allocator~\cite{pytorch-expandable}, break this assumption by reserving a large virtual-address range and mapping physical pages on demand to reduce fragmentation. Consequently, new physical pages may be mapped to an existing virtual address after the initial allocation and remain invisible to peer NPUs.

\name supports VMM API-based allocators by treating the runtime mapping as a lightweight incremental registration. It intercepts the mapping and asynchronously re-applies the shadow mapping for the newly mapped region. This avoids registering the entire virtual address range, which would be prohibitively expensive. If a freshly mapped region is accessed before the asynchronous UB mapping completes, \name still uses a synchronization barrier to ensure correctness, which is no worse than just-in-time registration.

Runtime remapping mainly appears in workloads with dynamic allocation patterns. For example, MoE serving with continuous batching~\cite{orca} can trigger remapping because both the active-token count and expert-routing distribution vary across batches~\cite{moe_variation}. To reduce the metadata exchange overhead, \name uses a fast broadcast path through CPU-side direct access to peer host DRAM, as described in \secref{sec:implementation}. Meanwhile, runtime remapping is triggered in less than 4\% of request batches in MoE serving. Even when it occurs, a gap of tens of milliseconds is still observed between the map operation and the first communication use, typically due to intervening computation, which is sufficient to hide UB mapping latency. As a result, \name supports VMM API-based allocators with less than 0.6\% end-to-end overhead.

\subsection{Memory Deregistration }
\label{sec:dereg}
Deallocation is handled symmetrically to allocation. Tensor frees that only return a sub-allocation block to the caching allocator do not change the underlying physical memory and trigger no action. Deregistration is required only when physical memory is actually released, such as \texttt{aclrtFree} on a physical segment or a VMM unmap that decommits pages. The key property enabling non-blocking deregistration is that UB address translation allows multiple virtual addresses to map to the same physical page. Therefore, each peer-side shadow mapping is only an independent alias of the released region. When an NPU frees a region, \name can immediately reclaim the local virtual address, while peer-side UB unmappings proceed asynchronously in the background. To avoid stale remote accesses, the released physical segment is not reused for new communication allocations until all corresponding UB mappings have been removed.

\section{Communication Library}
\label{sec:supernode_operator}


\subsection{Full-Mesh Programming Abstraction}
\label{sec:design-fullmesh}

\MyPara{Remote-slice execution model.}
As shown in \secref{sec:mot-operator}, synchronization overhead becomes increasingly expensive in the supernode. \name therefore adopts a full-mesh execution model for small-to-medium payloads, enabled by UB remote load/store. \name decomposes communication operators into a set of \emph{remote-slice transfers},  represented as
\[
    \langle peer, src, dst, bytes, op, flag \rangle 
\]
where \codeIn{peer} identifies the remote NPU, \codeIn{src} and \codeIn{dst} are virtual addresses in the shared address space, \codeIn{bytes} specifies the slice size, \codeIn{op} specifies the remote-memory action, and \codeIn{flag} encodes the optional dependency signal. Different communication operators are expressed by changing two fields. The \emph{slice map} defines which byte ranges are accessed by which peers, while the \emph{remote-memory action} specifies whether the transfer performs a load, a store, or an atomic instruction.

\MyPara{Minimal synchronization.}
The full-mesh execution model reduces synchronization at the algorithm level, and \name further minimizes synchronization in operator execution according to operator semantics. For data-movement operators such as AllGather and AllToAll, \name uses a pull mode, where each rank directly reads the remote slices and writes them into local output buffers, avoiding producer-consumer handshakes. For reduction-bearing operators such as AllReduce and ReduceScatter, remote operands are first loaded into the on-chip SRAM buffer and then accumulated into the local destination buffer. When multiple NPU cores update the same memory, \name uses the atomic instruction of the MTE during local write-back, so a store from SRAM to HBM becomes an atomic read-modify-write operation, which allows multiple NPU cores to accumulate partial results into the same destination without software locks. 

\MyPara{Unified kernel skeleton.}
The full-mesh abstraction separates operator semantics from backend execution. The operator frontend only generates the slice map and corresponding src/dst virtual addresses, while the backend schedules and executes remote-slice transfers. This separation allows backend optimizations to be applied below the unified programming abstraction, including workload-balanced NPU-core partitioning (\secref{sec:op-balance}) and NPU-driven SDMA offloading (\secref{sec:op-SDMA}).

\subsection{Workload-balanced NPU-core Partitioning}
\label{sec:op-balance}

A key challenge in implementing full-mesh operators with highly concurrent NPU cores is avoiding core-level long tails caused by imbalanced workload partitioning. The imbalance comes from two sources. \emph{(i)~Uneven traffic.} For routing- and shape-dependent operators, such as AllToAllv and MoE dispatch/combine, different peers may exchange different traffic volumes. \emph{(ii)~Non-uniform memory access.} UB's hierarchical topology introduces non-uniform latency and bandwidth, so the same traffic volume can incur different transfer latency depending on the relative locations of source and destination.

Naive partitioning policies, such as assigning each peer to a fixed NPU-core group, ignore both traffic skew and topology-dependent access cost. As a result, cores assigned to heavy or slow peer transfers finish much later, while other cores become idle at the local completion barrier.

\name models NPU-core partitioning as a minimum-makespan problem with a per-peer fairness constraint to avoid fan-out bursts and UB contention. For each peer \(p\), payload \(B_p\) is split into fine-grained transfer units. The unit size \(S_t\) is a tier-specific partitioning granularity chosen to minimize policy-generation overhead while keeping residual per-core imbalance bounded. Larger units reduce the cost of computing the partitioning policy, whereas smaller units provide finer-grained load balance. For a unit of size \(s \le S_t\), \name estimates its cycle cost as
\[
    \tau(s) = \alpha_t + \frac{s}{\beta_t},
\]
where \(\alpha_t\) is the measured access latency of tier \(t\), and \(\beta_t\) is the bandwidth. Thus, uneven traffic is captured by the number of units, while non-uniform UB access is captured by tier-dependent latency and bandwidth.

Given these units, \name assigns them to NPU cores and logical issue stripes. Each stripe represents one wave of concurrent remote-memory operations, where each core issues at most one transfer unit. Let \(x_{c,k,j}\in\{0,1\}\) indicate whether unit \(j\) is issued by core \(c\) in stripe \(k\), and let \(\tau_j\) be its estimated cost. The predicted load of core \(c\) is
\[
    L_c = \sum_{k}\sum_j x_{c,k,j}\tau_j .
\]
The primary objective is to minimize the predicted straggler:
\[
    \min \max_c L_c .
\]

To avoid bursty fan-out, \name further bounds the number of cores that simultaneously target the same peer. For peer \(p\), let \(\mathcal{J}_p\) denote all units targeting \(p\). The instantaneous fan-out to \(p\) in stripe \(k\) is
\[
    F_{k,p}=\sum_c\sum_{j\in\mathcal{J}_p}x_{c,k,j}.
\]
\name enforces a tier-specific cap:
\[
    F_{k,p}\le H_t,\quad \forall k,p ,
\]
where \(H_t\) is chosen according to the UB tier. This constraint spreads remote accesses across peers at each moment, rather than concentrating many cores on the same destination.

Since exact minimum-makespan scheduling is NP-hard, \name computes an approximate NPU-core partitioning policy with a longest-processing-time-first (LPT)-style list scheduler~\cite{graham-lpt}. Appendices \secref{app:partition-complexity} proves the NP-hardness of the partitioning problem and discusses the LPT-style approximation. The output is a per-core task list, and each NPU core simply follows the list to issue remote-memory instructions in order. Profiling results show that the relative completion-time gap between the fastest and slowest NPU cores is reduced from about 43\% to within 5\% after applying \name's partitioning policy (details in \secref{sec:abla-balance}).

The policy-generation overhead is less than 0.5\% of operator latency and can be reused across iterations for regular collectives with stable tensor shapes, as detailed in Appendices \secref{app:sched-overhead}. For dynamic MoE dispatch/combine, \name uses a hierarchical variant that preserves token-level placement while computing the NPU-core partitioning policy at expert-window and peer-window granularity, avoiding excessive token-level analysis overhead, as detailed in Appendices \secref{app:moe-sched}.

\subsection{NPU-Driven SDMA Offloading}
\label{sec:op-SDMA}

Saturating UB bandwidth with full-mesh operators is core-intensive. In the default path, data is moved through NPU cores' MTEs, keeping the cores occupied throughout the transfer. As shown in \secref{sec:mot-operator}, reaching 95\% of peak bus bandwidth requires using half of the NPU cores. This creates compute-resource contention under compute--communication overlap, such as FSDP parameter prefetch~\cite{fsdp} in LLM training and Two-micro-Batch Overlapping~\cite{TokenWeave} in LLM inference. To reduce NPU-core occupation, \name introduces an \emph{NPU-driven SDMA-offloaded} path, where data transfers are offloaded to the asynchronous SDMA engine, leaving NPU cores available for concurrent computation.

\begin{figure}[]
    \centering
    \begin{subfigure}[]{0.4\columnwidth}
      \centering
      \includegraphics[width=\linewidth]{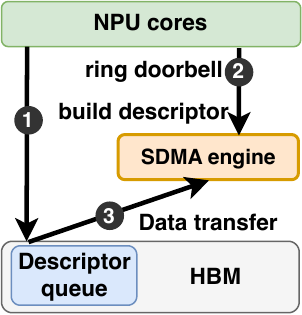}
      \phantomsubcaption
      \label{fig:sdma-offload-a}
    \end{subfigure}
    \hspace{0.04\columnwidth}%
    \begin{subfigure}[]{0.4\columnwidth}
      \centering
      \includegraphics[width=\linewidth]{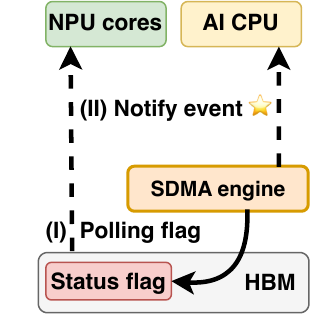}
      \phantomsubcaption
      \label{fig:sdma-offload-b}
    \end{subfigure}
    \caption{(a) Device-side SDMA issue path. (b) Two completion mechanisms: (I) NPU cores poll a status flag for in-kernel synchronization, and (II) the SDMA engine raises a notify event that a downstream or AI CPU waits on.}
    \label{fig:sdma-offload}
\end{figure}

The main challenge is issuing SDMA transfers with low control overhead. A conventional host-triggered path requires each transfer to pass through the CANN runtime and host control path, which is too expensive for communication operators with many fine-grained transfers. \name therefore issues SDMA entirely from the device side. During execution, NPU cores construct SDMA descriptors, submit them to per-core hardware queues concurrently, and ring the SDMA doorbell directly, as shown in Figure~\figref{fig:sdma-offload-a}.

\name supports two completion mechanisms depending on how the transferred data is consumed. For \emph{in-kernel consumption}, such as fused dispatch or combine, synchronization remains on the device side. After normal SDMA data descriptors, the sender appends a small tail descriptor that writes a status flag to the peer rank, and a few NPU cores poll local status flags to detect completion. For \emph{cross-kernel consumption}, such as collective communication followed by a separate compute kernel, \name uses SDMA notification. A notify record is appended to the same ordered SDMA queue, and the AI CPU waits for this notification before allowing the downstream to proceed. This mode introduces a kernel-level synchronization point, but fully releases NPU cores after descriptor submission, making it suitable for throughput-oriented overlap scenarios, as shown in Figure~\figref{fig:sdma-offload-b}.

SDMA offloading trades a small latency penalty for much lower NPU-core occupation. It is about 9\% slower than the MTE path due to descriptor-construction overhead, but reduces NPU-core occupation by over 95\%. This tradeoff benefits compute--communication overlap by leaving more NPU cores available to concurrent compute kernels (\secref{sec:ablation-sdma}).

\subsection{NPU-Side Profiler}
\label{sec:op-profiler}

Efficient communication-operator design on CM384 requires fine-grained visibility into intra-kernel behavior, including synchronization overhead and NPU-core imbalance. Existing profilers are insufficient for this purpose. System-level profilers, such as Nsight Systems~\cite{nsight_system} and MindStudio~\cite{mindstudio}, mainly provide inter-kernel timelines with coarse-grained attribution. Hardware-metric profilers, such as Nsight Compute~\cite{nsight_compute}, expose hardware information but do not directly attribute latency to developer-defined communication stages.

To address this gap, \name provides a lightweight NPU-side profiler that records per-stage timing inside communication kernels. The profiler inserts inline probes at protocol-stage boundaries and uses the on-core cycle counter to timestamp developer-defined stages. Each NPU core writes events to its own slice of a host-pinned trace buffer, and then decodes the trace into per-core timelines for inspecting per-stage latency.
The profiler supports both manual and automatic instrumentation. In manual mode, developers insert named begin/end probes around semantic regions. In automatic mode, inspired by Neutrino's programmable assembly-level GPU probing~\cite{neutrino}, \name injects probes into compiled kernels without source-code modification. The collected traces expose per-core and per-stage latency, helping identify synchronization tails, topology-dependent slow paths, and excessive NPU-core usage, which in turn guides the optimization of efficient communication operators.

\section{Implementation}
\label{sec:implementation}

\name is implemented on top of the CANN stack~\cite{CANN}. The implementation contains about 21K lines of Ascend C/C++ and 5K lines of Python, and is integrated with PyTorch~\cite{PyTorch}, SGLang~\cite{sglang}, TorchTitan~\cite{torchtitan}, and TorchRec~\cite{torch_rec}.

\MyPara{Interception layer.}
\name makes registration-on-allocation transparent by intercepting framework memory APIs. It interposes CANN allocation and mapping entry points, including \texttt{aclrtMalloc} and \texttt{aclrtMapMem}. The hook is triggered only by physical-memory events, such as newly created caching-allocator segments or VMM page mapping.

\MyPara{CPU-side direct DRAM access.}
\name enables host CPUs to access peer DRAM over UB with regular load/store instructions, which is not supported by the default supernode configuration. \name extends the global UB physical-address map with a remote-DRAM range, configures BIOS routing through CPU-connected UB switch planes, and installs address-decode and translation entries in UB switches. Remote-DRAM UB addresses are then mapped into the accessing process's CPU page table, allowing CPU cores to access peer host DRAM through ordinary virtual addresses. This direct path reduces metadata-broadcast latency from 60~\us to 7--8~\us compared with the NPU-relay path.

\MyPara{Generality on NVIDIA GPUs.} \name relies on a scale-up fabric with a global address space to support registration-on-allocation, and similar capabilities are available on NVIDIA supernodes such as GB200 NVL72~\cite{gb200nvl72}. To evaluate this portability, a prototype integrates \name with the CUDA runtime and NCCL on NVIDIA GPUs. The results show that \name can also move user-buffer registration off the communication critical path and improve NCCL-based communication, as detailed in \secref{sec:ablation-nvidia}.

\section{Evaluation}
\label{sec:evaluation}


This section evaluates \name at both the operator level (\secref{sec:eval-microbench}) and the end-to-end application level (\secref{sec:eval-end2end}). All experiments are conducted on a CM384 supernode.

\subsection{Microbenchmark}
\label{sec:eval-microbench}

\MyPara{Settings and metrics.}
This section evaluates two operator families at the 32-rank scale, using 32 NPU dies across two nodes. For collectives, the per-rank payload is swept from 1~MiB to 1~GiB. For MoE dispatch/combine, the evaluation follows prior settings~\cite{deepep,swiftep}: 4096 tokens per batch, hidden size 7168, top-8 experts, and EP=32, covering both high-throughput (HT) and low-latency (LL) modes. Across all experiments, performance is reported as bus bandwidth (Bus BW) in GB/s, following the standard NCCL convention~\cite{bus_bandwidth_NCCL}.

\MyPara{Baselines.}
For collectives, the baselines are HCCL~\cite{hccl}, the production collective communication library for Ascend NPUs, which incorporates state-of-the-art optimizations such as NVSHMEM-style symmetric memory~\cite{nvshmem}, and HCCL-zerocopy~\cite{hccl-zerocopy}, which enables user-buffer direct communication. For MoE dispatch/combine, the baselines include DeepEP v2~\cite{deepep} over CX7 400~Gb/s InfiniBand RDMA, a state-of-the-art expert-parallel communication library, and the Ascend implementation of DeepEP~\cite{DeepEP-Ascend} enabled by HCCL, denoted as CANN EP, together with its zerocopy variant.

\begin{figure}[t]
\centering
\includegraphics[width=\columnwidth]{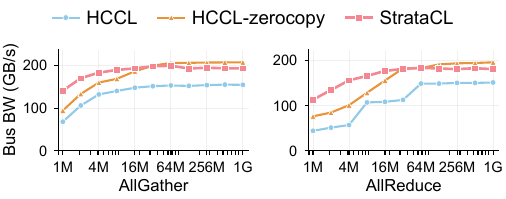}
\caption{Bus bandwidth of AllGather and AllReduce.}
\label{fig:eval-microbench}
\end{figure}

\begin{table}[t]
\centering
\footnotesize
\caption{MoE dispatch/combine Bus BW (GB/s) at EP=32. }
\label{tab:eval-ep}
\begin{tabular}{ccccc}
\toprule
 & \makecell{\textbf{DeepEP}}
 & \makecell{\textbf{CANN EP}}
 & \makecell{\textbf{CANN EP}\\\textbf{zerocopy}}
 & \makecell{\textbf{\name}} \\
\midrule
Dispatch (HT) & 61 & 98 & 106 & \textbf{130} \\
Combine (HT)  & 61 & 85  & 93 & \textbf{121} \\
Dispatch (LL) & 50 & 61  & 82  & \textbf{107} \\
Combine (LL)  & 55 & 68  & 79  & \textbf{108} \\
\bottomrule
\end{tabular}
\end{table}

\MyPara{Collectives.}
Figure~\figref{fig:eval-microbench} reports AllGather and AllReduce performance. HCCL-zerocopy improves over HCCL by ${\sim}1.3\times$ on average, as user-buffer direct communication removes redundant copies between user tensors and internal communication buffers. \name further outperforms HCCL-zerocopy in the small-to-medium payload regime, achieving up to $1.6\times$ higher bus bandwidth. The gains come from two optimizations. First, full-mesh execution avoids the multi-step synchronization overhead. Second, workload-balanced NPU-core partitioning mitigates core-level long tails caused by traffic skew and non-uniform memory access. For large payloads, HCCL-zerocopy can slightly outperform \name, with ${\sim}6\%$ higher bus bandwidth on average, because full-mesh execution introduces higher fan-out and network contention when too many remote accesses are issued concurrently. This limitation can be addressed by a workload-aware operator selection policy that switches to multi-step algorithms for large payloads. Extended results for the remaining collectives are reported in Appendices~\secref{sec:appendix_microbenchmark}.

\MyPara{MoE dispatch/combine.}
Table~\ref{tab:eval-ep} reports dispatch/combine performance at EP=32. CANN EP outperforms RDMA-based DeepEP by $1.4\times$ on average, because CM384's UB fabric provides higher physical bandwidth than CX7 400~Gb/s InfiniBand and avoids the NIC-forwarding path used by RDMA-based expert-parallel communication. Enabling zerocopy further improves CANN EP by eliminating staging copies between user buffers and communication buffers, increasing bus bandwidth by $8.8\%$ on average in HT mode and $25.3\%$ in LL mode. \name achieves the highest bus bandwidth in all cases, outperforming CANN EP zerocopy by $22.6\%$--$36.7\%$. These gains come from \name's workload-balanced NPU-core partitioning, which accounts for expert-routing imbalance and mitigates core-level long tails.

\subsection{End-to-end Performance}
\label{sec:eval-end2end}

\begin{table}[t]
\centering
\footnotesize
\caption{End-to-end workload settings.}
\label{tab:e2e-settings}
\begin{tabular}{cccc}
\toprule
\textbf{Workload} & \textbf{Model} & \textbf{NPU dies} & \textbf{Parallelism} \\
\midrule
LLM inference      & DS V4 Flash    & 192 & DP+EP \\
LLM training       & DS V3.2 671B   & 512 & FSDP+TP+EP \\
Recsys training    & DLRM                 & 128 & DP+MP (TW emb.) \\
\bottomrule
\end{tabular}
\end{table}

\MyPara{Workloads.}
This section evaluates three representative production workloads, including LLM inference, LLM training, and recommendation-system (Recsys) training (Table~\figref{tab:e2e-settings}).

\underline{\emph{LLM inference:}}
DeepSeek (DS) V4 Flash~\cite{deepseek_v4} is evaluated with SGLang~\cite{sglang} under a disaggregated serving setting~\cite{distserve,splitwise}. Each prefill instance uses DP=32 for attention layers and EP=32 for MoE layers, while each decode instance uses EP=16 and DP=16. The serving stack further adopts state-of-the-art scheduling mechanisms, including Two-micro-Batch Overlapping (TBO)~\cite{profile-data,TokenWeave} and Multi-Token Prediction (MTP)~\cite{mtp}. The deployment uses four prefill instances on eight nodes and four decode instances on four nodes, totaling 192 NPU dies. The request length distribution follows the Splitwise conversation dataset~\cite{splitwise}, whose average input-to-output length ratio is approximately 8:1.

\underline{\emph{LLM training:}}
DeepSeek V3.2 671B~\cite{deepseek_v3_2} is evaluated with TorchTitan~\cite{torchtitan} on 512 NPU dies. The training run uses hybrid parallelism with FSDP=128, TP=4, and EP=64. The global batch size is 512, and the sequence length is 4096. This workload includes both common collectives and MoE-specific communication: FSDP introduces parameter AllGather and gradient ReduceScatter, TP invokes AllReduce, and MoE layers invoke dispatch/combine for expert routing. FSDP parameter prefetch is enabled, allowing parameter AllGather to overlap with adjacent-layer computation.

\underline{\emph{Recsys training:}}
DLRM~\cite{DLRM} is evaluated with TorchRec~\cite{torch_rec} on 128 NPU dies using a standard hybrid-parallel deployment. The embedding tables are table-wise sharded across dies, while the dense MLPs are replicated with data parallelism. The Criteo Terabyte dataset~\cite{criteo_TB} drives embedding accesses with a batch size of 1024 and a 7~TiB embedding table. This workload involves AllToAll and AllReduce. AllToAll routes sparse features and exchanges pooled embeddings across embedding-table owners, while AllReduce synchronizes dense-MLP gradients across data-parallel replicas.

\MyPara{Metrics.}
For LLM inference, performance is measured under different request rates. The primary metric is normalized latency, computed as end-to-end request latency divided by the number of generated tokens. P99 time-to-first-token (TTFT) and P99 time-per-output-token (TPOT) are also reported. For training, iteration time is used as the primary metric.

\MyPara{Baselines.}
\name is compared with two baselines, HCCL and HCCL-zerocopy. Since the native HCCL-zerocopy interface is not integrated with existing production frameworks, HCCL-zerocopy is implemented as a practical and optimistic baseline using pool-based registration. Specifically, the whole HBM is pre-registered as communication-accessible memory, and user tensors are allocated from this pool, avoiding just-in-time registration on the critical path.

\begin{figure}[t]
\centering
\begin{subcaptiongroup}
\includegraphics[width=\columnwidth]{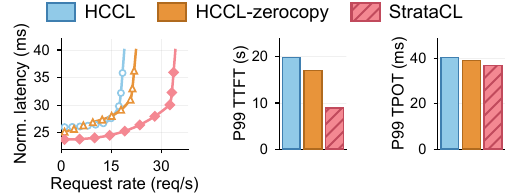}
\phantomsubcaption\label{fig:eval-llm-a}
\phantomsubcaption\label{fig:eval-llm-b}
\end{subcaptiongroup}
\caption{LLM inference performance. 
(a) Normalized latency vs. request rate. 
(b) P99 TTFT and P99 TPOT at 15~req/s.} 
\label{fig:eval-llm}
\end{figure}

\MyPara{LLM inference.}
Figure~\figref{fig:eval-llm-a} reports normalized latency under different request rates. HCCL-zerocopy improves serving throughput over HCCL by only $1.2\times$. Although user-buffer direct communication reduces communication latency, its pre-registered memory pool disables PyTorch's expandable-segment allocator and adds 1.6~GiB of fragmentation. This forces the serving batch size to drop by 3 to avoid out-of-memory errors, partially offsetting the operator-level benefit.

\name avoids this problem by registration-on-allocation, preserving PyTorch's expandable-segment mechanism and maintaining the same serving batch size as HCCL. As a result, \name improves inference throughput by $1.9\times$ over HCCL and $1.6\times$ over HCCL-zerocopy. The gain comes from workload-balanced NPU-core partitioning and SDMA offloading, which prevent communication kernels from competing with compute kernels for NPU cores under TBO. 

Figure~\figref{fig:eval-llm-b} further reports P99 TTFT and P99 TPOT at 15~req/s. Compared with HCCL, \name reduces P99 TTFT and P99 TPOT by $2.2\times$ and $1.1\times$, respectively. These results show that \name improves both peak serving throughput and tail latency under high-load serving conditions.

\begin{figure}[t]
\centering
\begin{subcaptiongroup}
\includegraphics[width=\columnwidth]{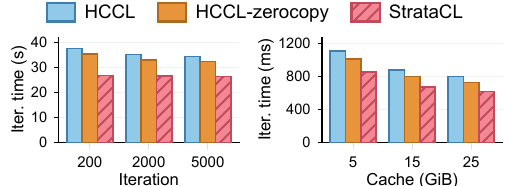}
\phantomsubcaption\label{fig:eval-train-a}
\phantomsubcaption\label{fig:eval-train-b}
\end{subcaptiongroup}
\caption{ 
(a) LLM training iteration time at different training iterations. 
(b) Recsys training iteration time under different embedding cache sizes. Cache size here refers to the memory allocated to the embedding table for each NPU.}
\label{fig:eval-train}
\end{figure}

\MyPara{LLM training.}
Figure~\figref{fig:eval-train-a} reports LLM training iteration time. HCCL-zerocopy provides only a modest improvement over HCCL, reducing iteration time by 6\%, because training imposes stronger and longer-lived memory pressure than inference, making memory fragmentation more severe, which introduces about 3~GiB of additional fragmentation. In contrast, \name reduces iteration time by 18\%--24\% over HCCL-zerocopy. This gain comes from registration-on-allocation, which avoids pool-based fragmentation, and SDMA offloading, which reduces NPU-core contention between FSDP prefetch communication and concurrent computation. The benefit becomes smaller in later iterations because gate layers gradually improve expert load balance~\cite{sparsely_gated_moe}, making traffic distribution more uniform and reducing the opportunity for workload-balanced NPU-core partitioning.

\MyPara{Recsys training.}
Figure~\figref{fig:eval-train-b} reports the Recsys training iteration time. \name reduces iteration time by ${\sim}23\%$ over HCCL and ${\sim}16\%$ over HCCL-zerocopy, demonstrating its effectiveness for sparse embedding communication. The communication benefit becomes more pronounced when the embedding cache is smaller, or the per-worker batch size is larger, because both cases increase embedding-cache misses and trigger more remote embedding transfers, making communication a larger fraction of iteration time.

\section{Ablation Study}

\subsection{Performance Breakdown}
\label{sec:ablation-breakdown}

\begin{figure}[]
\centering
\begin{subcaptiongroup}
\includegraphics[width=\columnwidth]{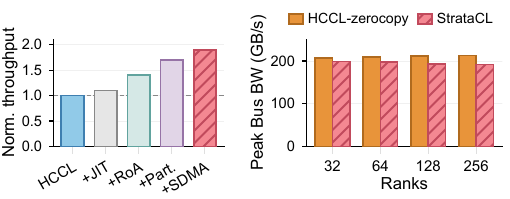}
\phantomsubcaption\label{fig:eval-breakdown}
\phantomsubcaption\label{fig:eval-scal}
\end{subcaptiongroup}
\caption{(a) Performance breakdown of \name's LLM inference throughput (JIT: just-in-time user-buffer registration, RoA: registration-on-allocation, Part.: workload-balanced NPU-core partitioning). (b) Peak AllGather bus bandwidth.}
\label{fig:eval-breakdown-scal}
\end{figure}

Figure~\figref{fig:eval-breakdown} decomposes \name's end-to-end LLM inference gain by enabling one technique at a time on top of HCCL. Registering user buffers just-in-time yields only $1.1\times$, because per-buffer registration latency consumes most of the zero-copy benefit. Registration-on-allocation alone improves throughput to $1.4\times$, because it enables user-buffer direct communication without the registration overhead on the critical path. Adding workload-balanced NPU-core partitioning raises the gain to $1.7\times$ by reducing core-level long tails in collectives and dispatch/combine operators. Finally, SDMA offloading releases NPU cores during data transfers, lifting the throughput to $1.9\times$. This is because communication no longer competes with computation for NPU cores.

\subsection{Communication Operator Scalability}
\label{sec:ablation-scal}

Figure~\figref{fig:eval-scal} compares the peak AllGather bus bandwidth of \name and HCCL-zerocopy as the communicator scales from 32 to 256 ranks. HCCL-zerocopy sustains a slightly higher peak bandwidth with its multi-step ring algorithm, while \name's full-mesh execution becomes modestly lower at larger scales due to increasing fan-out and network contention from concurrent remote accesses. However, this gap grows sublinearly with rank count and remains within 10\% even at 256 ranks. The remaining large-payload peak-bandwidth gap can be further reduced by a workload-aware operator selection policy that switches to multi-step algorithms for large payloads.

\subsection{Generality on NVIDIA GPUs}
\label{sec:ablation-nvidia}

\begin{figure}[t]
\centering
\begin{subcaptiongroup}
\includegraphics[width=\columnwidth]{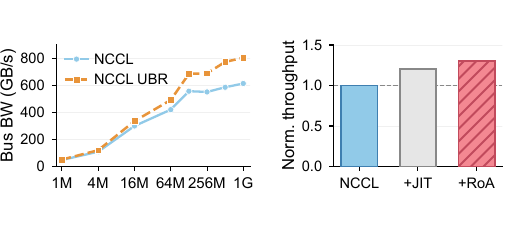}
\phantomsubcaption\label{fig:eval-nvidia-a}
\phantomsubcaption\label{fig:eval-nvidia-b}
\end{subcaptiongroup}
\caption{(a) AllGather Bus BW of NCCL vs.\ NCCL UBR. (b) DeepSeek V4 Flash~\cite{deepseek_v4} serving throughput with SGLang.}
\label{fig:eval-nvidia}
\end{figure}

To validate the generality, a prototype is implemented on NVIDIA GPUs and evaluated on an NVIDIA DGX B200 server with NVLink~5.0, serving as a smaller NVLink-domain proxy for NVIDIA supernodes such as GB200 NVL72~\cite{gb200nvl72}. Figure~\figref{fig:eval-nvidia-a} reports 8-rank AllGather performance with NCCL and NCCL User Buffer Registration (UBR)~\cite{nccl-ubr}. NCCL UBR removes staging copies and improves bus bandwidth over baseline NCCL across all payloads, with an average gain of $1.2\times$.

Figure~\figref{fig:eval-nvidia-b} reports end-to-end serving throughput for DeepSeek V4 Flash~\cite{deepseek_v4} with SGLang~\cite{sglang}. NCCL UBR with just-in-time registration improves throughput to $1.2\times$ by enabling user-buffer direct communication, while NCCL with registration-on-allocation further reaches $1.3\times$ by moving buffer registration off the communication critical path. The gain is moderate in this 8-GPU setting because the cost of just-in-time registration is not strongly amplified at a small scale. At larger NVLink-domain scales, the benefit of registration-on-allocation is expected to increase because registration overhead grows with the number of peers, consistent with the scaling behavior observed on CM384. These results show that the core idea of registration-on-allocation is not specific to CM384 and can also benefit NCCL-based communication on NVIDIA scale-up fabrics.

\subsection{Allocation-to-communication gap statistics}
\label{sec:ablation-gap}

\begin{figure}[t]
\centering
\begin{subfigure}[b]{0.44\columnwidth}
  \centering
  \includegraphics[width=\linewidth]{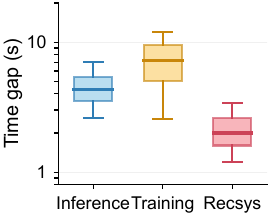}
  \phantomsubcaption
  \label{fig:ablation-gap}
\end{subfigure}
\hfill
\begin{subfigure}[b]{0.54\columnwidth}
  \centering
  \includegraphics[width=\linewidth]{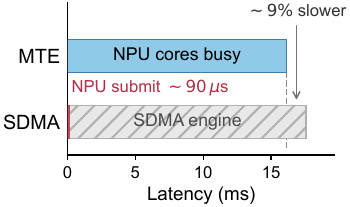}
  \phantomsubcaption
  \label{fig:eval-sdma}
\end{subfigure}
\caption{(a) Allocation-to-communication gap. (b) NPU-core occupation time under SDMA offloading.}
\label{fig:ablation}
\end{figure}

This section evaluates whether registration-on-allocation can hide remote-access registration from the communication critical path (\secref{sec:reg-on-alloc}). Across the three production workloads, \name profiles each communication-operator invocation and traces the accessed buffers back to their physical allocation time using PyTorch's caching-allocator memory history~\cite{pytorch_memhistroy}. As shown in Figure~\figref{fig:ablation-gap}, even the minimum allocation-to-communication gap is several seconds, orders of magnitude larger than the microsecond- to millisecond-scale cost of remote-access registration. This interval mainly consists of model warm-up activities, such as Graph capture~\cite{acl_graph} and KV-cache~\cite{kv_cache} allocation in LLM inference or model weight loading in LLM training. These results confirm that registration-on-allocation can remove remote-access registration from the communication critical path.

\subsection{NPU-driven SDMA Offloading}
\label{sec:ablation-sdma}

This section evaluates whether SDMA offloading reduces NPU-core occupation (\secref{sec:op-SDMA}). Using the NPU-side profiler, a 32-rank 128~MiB AllGather is profiled under two paths. In the MTE path, NPU cores directly perform data movement. In the SDMA-offloaded path, NPU cores only submit descriptors, while the SDMA engine transfers data asynchronously. 

As shown in Figure~\figref{fig:eval-sdma}, the MTE path keeps NPU cores busy for almost the entire transfer, while the SDMA-offloaded path uses them only for descriptor submission, reducing NPU-core occupation by over 95\%. This comes with a modest 9\% latency slowdown due to descriptor construction and doorbell submission. This tradeoff is favorable because released NPU cores reduce contention with concurrent compute kernels and improve end-to-end throughput.

\subsection{NPU-Core Workload Balance}
\label{sec:abla-balance}

\begin{figure}[t]
    \centering
    \includegraphics[width=0.82\columnwidth]{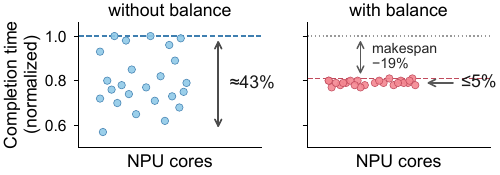}
    \caption{Per-NPU-core completion time statistics.}
    \label{fig:eval-balance}
\end{figure}

This section evaluates whether workload-balanced NPU-core partitioning reduces core-level long tails (\secref{sec:op-balance}). Using the NPU-side profiler, per-core completion time is collected on a 32-rank 128~MiB AllGather under naive equal-core partitioning and \name's workload-balanced partitioning.

As shown in Figure~\figref{fig:eval-balance}, naive partitioning creates a clear tail because cores assigned to slower or heavier peer transfers finish much later, leaving faster cores idle. The fastest-to-slowest completion-time gap reaches about 43\%. By assigning transfer units according to modeled tier cost and bounding instantaneous fan-out, \name reduces the gap to within 5\% and lowers the makespan by 19\%.

\section{Related Work}
\label{sec:related}

\MyPara{Collective communication libraries.}
Vendor stacks such as NCCL, HCCL, and RCCL provide collectives for GPU and NPU~\cite{nccl,hccl,rccl}, while NVSHMEM, rocSHMEM, and CANN SHMEM expose PGAS-style remote memory operations to device kernels~\cite{nvshmem,roc_shmem,aclshmem,PGAS}. Programmable libraries such as MSCCL++ and MSCCLang provide algorithm- or schedule-level primitives for customized optimization~\cite{mscclpp,mscclang}, and TACCL synthesizes topology-aware collective algorithms from communication sketches~\cite{taccl}. NCCL UBR~\cite{nccl-ubr} and HCCL-zerocopy~\cite{hccl-zerocopy} reduce redundant copies through user-buffer direct communication, but still rely on just-in-time registration. \name advances beyond these mechanisms with registration-on-allocation, which registers physical allocations asynchronously, remains compatible with VMM API-based allocators, and keeps user-buffer direct communication transparent to production frameworks.

\MyPara{Expert-parallel communication.}
MoE models rely on expert-parallel dispatch and combine, which have become major communication bottlenecks in large-scale training and serving. DeepEP, Tutel, FlashMoE, and MegaScale-MoE optimize token shuffling, fused dispatch/combine kernels, expert routing, and communication--computation overlap~\cite{deepep,Tutel,flashmoe,megascale-MoE}. SwiftEP further reduces staging overhead through buffer fusion~\cite{swiftep}. These techniques mainly optimize how MoE data movement is packed, scheduled, fused, or overlapped, while remote-accessible buffers are still typically prepared through pre-allocated memory pools. \name instead removes registration from the communication critical path through registration-on-allocation and optimizes dispatch/combine considering supernode architectural features.

\MyPara{Communication in distributed ML systems.}
Communication efficiency is a first-class concern across the distributed-ML stack. Alibaba HPN tailors datacenter networks to LLM traffic~\cite{hpn}, Vela provides virtualized GPU-direct RoCE fabrics~\cite{vela}, and HostNet redesigns host networking by separating a zero-copy data path from a flexible control path~\cite{hostnet}. MegaScale co-designs parallelism and communication to scale training beyond 10{,}000 GPUs~\cite{megascale}, Alpa automatically derives parallelization plans~\cite{alpa}, and TopoOpt co-optimizes network topology and parallelization strategy~\cite{topoopt}. What-if analysis diagnoses communication stragglers~\cite{stragglers}, while PipeMorph adapts pipeline schedules to tolerate communication jitter~\cite{pipemorph}. These systems optimize at the cluster and framework layers, which are orthogonal to \name's focus on communication-library optimization.

\MyPara{Communication--computation overlap.}
Another line of work hides communication latency behind computation. T3 transparently tracks and triggers collectives to overlap them with dependent computation~\cite{T3}, while TokenWeave splits token batches so that one wave's communication overlaps with another wave's computation~\cite{TokenWeave}. Lagom co-tunes communication and computation to maximize overlap in distributed LLM training~\cite{lagom}. Recent systems also offload communication to dedicated engines. ARK lets GPU kernels drive DMA engines without CPU intervention~\cite{ark}, while ConCCL and DMA Collectives offload concurrent collectives to GPU DMA/copy engines~\cite{conccl,dmacollectives}. \name achieves a similar goal in the supernode by integrating NPU-driven SDMA offloading. This releases NPU cores and avoids compute-resource contention in production overlap scenarios.

\section{Conclusion}
This paper presents \name, a zero-redundancy and fabric-native communication library for production supernodes. Through the registration-on-allocation mechanism, \name removes staging copies and keeps remote-access registration off the critical path while remaining transparent to applications. It further reduces synchronization overhead, core-level long tails, and core contention through full-mesh execution, workload-balanced NPU-core partitioning, and SDMA offloading. Evaluation on CM384 shows that \name improves collective and MoE communication and delivers significant end-to-end gains across LLM inference, LLM training, and Recsys training, demonstrating its practicality for accelerating distributed AI applications on production supernodes.

\bibliographystyle{ACM-Reference-Format}
\bibliography{biblio}

\appendix

\clearpage

\section*{APPENDICES}

\section{Extended Microbenchmark}
\label{sec:appendix_microbenchmark}

\begin{figure}[t]
\centering
\includegraphics[width=\columnwidth]{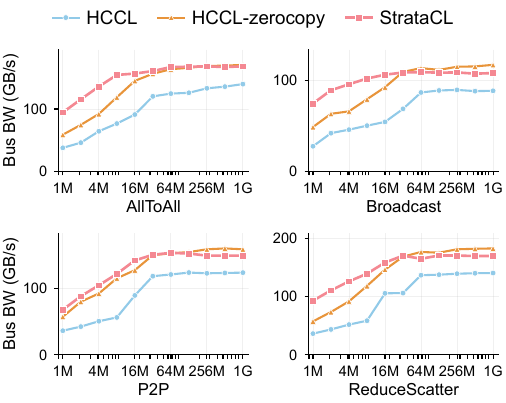}
\caption{Extended collective microbenchmarks.}
\label{fig:eval-microbench-appendix}
\end{figure}

Figure~\figref{fig:eval-microbench-appendix} reports the remaining four collective primitives, namely AllToAll, Broadcast, P2P, and ReduceScatter, under the same setup as \secref{sec:eval-microbench}. The results show a consistent trend with AllGather and AllReduce. HCCL-zerocopy improves over HCCL by eliminating redundant staging copies, while \name further improves performance in the small-to-medium payload regime. For P2P transfer, \name achieves performance close to HCCL-zerocopy, because both paths reduce to a single direct peer-to-peer data transfer with little algorithmic synchronization or workload-partitioning opportunity.

\section{Complexity of NPU-Core Partitioning}
\label{app:partition-complexity}

This appendix analyzes the complexity of the NPU-core partitioning problem in \secref{sec:op-balance}. Given transfer units \(\mathcal{J}\), \(C\) NPU cores, estimated unit costs \(\tau_j\), peer mapping \(p(j)\), and tier-specific fan-out caps \(H_t\), the decision problem asks whether the units can be assigned to cores and issue stripes such that the predicted makespan is at most \(T\) while satisfying the per-peer fan-out cap:
\[
    \max_c \sum_{k}\sum_j x_{c,k,j}\tau_j \le T,
    \qquad
    \sum_c\sum_{j\in\mathcal{J}_p} x_{c,k,j} \le H_{t(p)},\quad \forall k,p ,
\]
where \(x_{c,k,j}\in\{0,1\}\) indicates that unit \(j\) is issued by core \(c\) in stripe \(k\), each unit is assigned exactly once, each core issues at most one unit per stripe, and \(\mathcal{J}_p=\{j \mid p(j)=p\}\) denotes the units targeting peer \(p\), whose UB tier is \(t(p)\).

\MyPara{NP-hardness by reduction.}
The decision version of NPU-core partitioning is NP-hard by reduction from the classical minimum-makespan scheduling problem on identical machines, \(P||C_{\max}\). Given an instance of \(P||C_{\max}\) with \(C\) machines, jobs \(\mathcal{J}\), processing times \(a_j\), and target makespan \(T\), construct an NPU-core partitioning instance with \(C\) cores and one transfer unit per job. Set each unit cost to \(\tau_j=a_j\), place all units in the same UB tier and peer group, and set the fan-out cap to \(H_t=C\). Since each core issues at most one transfer unit per stripe, any stripe contains at most \(C\) units in total, so the fan-out cap is never binding. The problem then reduces exactly to assigning jobs to \(C\) cores while minimizing the maximum per-core load. Thus, the constructed NPU-core partitioning instance has a feasible solution with makespan at most \(T\) if and only if the original \(P||C_{\max}\) instance has a feasible schedule with makespan at most \(T\). Since \(P||C_{\max}\) is NP-hard, NPU-core partitioning is also NP-hard.

\MyPara{LPT-style approximate partitioning.}
Because exact solving requires expensive combinatorial optimization, \name uses a lightweight longest-processing-time-first (LPT)-style list scheduler~\cite{graham-lpt}.
Without the fan-out constraint, this reduces to classical LPT scheduling, whose makespan satisfies
\[
    C_{\max}^{\mathrm{LPT}}
    \le
    \left(\frac{4}{3}-\frac{1}{3C}\right) C_{\max}^{*},
\]
where \(C_{\max}^{*}\) is the optimal makespan. In \name, the stripe-level fan-out cap makes the problem a constrained variant, so the classical bound does not directly apply. 

\section{Workload Partitioning Overhead Analysis}
\label{app:sched-overhead}

This appendix analyzes the overhead of computing the workload partitioning policy used in \secref{sec:op-balance}. Let \(U=\sum_p \lceil B_p/S_{t(p)} \rceil\) be the total number of atomic transfer units, and let \(C\) be the number of NPU cores used for communication. To generate the partitioning policy, \name first sorts transfer units by their estimated cycle cost and then assigns each unit to the least-loaded feasible NPU core under the stripe-level fan-out cap. Therefore, the theoretical time complexity is
\[
    O(U\log U + U\log C),
\]
where \(O(U\log U)\) comes from sorting transfer units and \(O(U\log C)\) comes from maintaining the least-loaded core during LPT-style list assignment.

In practice, this overhead is small. For the 32-rank 128~MiB AllGather used in the evaluation, each rank computes the partitioning policy over 24 communication NPU cores. Generating the policy takes about 40~\us, which is less than 0.5\% of the operator end-to-end latency. This cost can also overlap with metadata exchange, further reducing its exposure on the critical path. Moreover, the policy is reusable: for regular collectives with stable tensor shapes, \name caches the workload partitioning policy, so later invocations reuse the cached per-core task queues without recomputation.

\section{Hierarchical Workload Partitioning for MoE Dispatch/Combine}
\label{app:moe-sched}

MoE dispatch and combine have dynamic communication patterns because expert routing changes across batches. A naive design could treat every routed token as an independent unit when computing the workload partitioning policy. This would make the policy-generation overhead scale with the number of routed tokens, which is especially expensive during prefill when the token count can be large.

\name avoids this overhead with a hierarchical workload partitioning design. The first level is \emph{token placement}. For each batch, the MoE routing path already computes the destination expert, per-expert counts, prefix-sum offsets, and token-local offsets required by dispatch and combine. \name reuses this metadata and does not introduce an additional token-level optimization pass. The second level is \emph{expert-window aggregation}. Tokens with the same destination peer and expert are grouped into a contiguous expert window, identified by a base offset and token-local indices. This preserves token-level placement while exposing compact communication ranges. The third level is \emph{peer-window partitioning}. \name aggregates expert windows by peer to obtain the peer payload \(B_p\), UB tier \(t(p)\), and transfer-unit count \(n_p=\lceil B_p/S_{t(p)}\rceil\). The NPU-core partitioning policy is then computed over peer-window ranges under the same cycle-cost model and stripe-level fan-out cap, instead of materializing every routed token as a separate scheduling unit.

This hierarchy separates token placement from NPU-core workload partitioning. Token placement remains token-level for correctness, while workload partitioning is performed at expert-window and peer-window granularity. As a result, the input size for policy generation is bounded by the number of active expert windows and peers, rather than the number of routed tokens. This keeps the overhead small even for long-prefill MoE batches, while still adapting to dynamic routing skew.


\end{document}